# High-quality metalens enables minimally invasive CFB endoscopy

**Ruixiang Song[1,2], Xutong Lu[1,2], Xiyao Song[1], Shuaihong Qi[1,2], Feng Wang[1,2], Jiaqi Cui[1,2], Zhangyuan Chen[1,2] and Yanping Li[1,2,*]**

[1] *State Key Laboratory of Photonics and Communications, School of Electronics, Peking University, Beijing 100871, China.*

[2] *Frontiers Science Center for Nano-optoelectronics, Peking University, Beijing 100871, China.*

* *liyp@pku.edu.cn*

**Abstract:** Metalenses, owing to their ultra-thin planar structures, present a promising solution for reducing endoscopic invasiveness. However, achieving high-quality imaging with minimal invasiveness (short focal length of metalens) remains a critical challenge. This paper presents a deep learning assisted metalens with a 1mm focal length tailored for coherent fiber bundle, constituting the least invasive metalens-CFB system reported to date. To overcome the increased chromatic dispersion and aberrations associated with high NA metalens, we compensate for lateral etching at the base of the nanopillars by adjusting the thickness of a sacrificial hard mask. This approach enables the fabrication of nanopillars with small cross sections and high aspect ratios, featuring nearly vertical sidewalls (~90°), thereby enhancing the phase accuracy of the metalens. Experimental validation using the metalens-CFB system demonstrates that the metalens achieves an expanded field of view of 48.3° and a depth of field exceeding 125 mm. This work establishes a new paradigm for ultra-minimally invasive endoscopic imaging.

## 1. Introduction

As a medical imaging device for direct visualization of internal organs [1], the endoscope typically comprises three key components: an image transmission medium, a backend image processing module, and a frontend auxiliary imaging component. It allows physicians to observe narrow and inaccessible regions of the human body, such as blood vessels [2-4], the brain, and the spinal cord [5, 6], with diameters of only a few millimeters. In such scenarios, an endoscope with a wide field of view (FOV) and large depth of field (DOF) can significantly improve diagnostic efficiency. However, conventional endoscopic imaging systems often involve trade-offs between spatial resolution, FOV, DOF, and invasiveness. Various auxiliary devices at distal end have been proposed for high-resolution imaging. For instance, integrating piezoelectric actuator structure [7] into the catheter can significantly enhance both resolution and FOV, but this improvement is achieved at the cost of increased probe diameter, leading to greater invasiveness and patient discomfort. GRIN lenses [8] offer a compact, easily integrated alternative at the distal end, but their intrinsic aberrations limit the FOV and their short working distance restricts imaging near the lens surface. Using a metalens with intrinsic chromatic aberration in combination with a coherent fiber bundle (CFB) for quantitative phase imaging can ease metalens design complexity, but slight variations in fiber core length severely degrade imaging quality [9].

In recent years, metalenses have emerged as a promising solution to address the trade-offs in CFB endoscopic imaging systems discussed above. Composed of periodically arranged nanostructures, metalenses enable subwavelength-resolution light manipulation [10-14]. Through precise light manipulation, metalenses avoid spherical aberration [15-17] and thus improve image resolution, whereas conventional aspherical lenses require costly and time-consuming processes. An additional advantage of metalenses is their ultrathin, planar architecture, with thickness determined solely by the height of the nanopillars and typically on the order of the incident wavelength. This structural compactness greatly reduces invasiveness. Furthermore，metalenses can significantly expand the FOV of CFB endoscopy, where the FOV is inherently limited due to the narrow acceptance angle of individual fiber cores, typically around 8° [18]. Integrating a metalens at the distal end of the CFB can significantly enlarge the FOV. Moreover, the intrinsic large DOF of metalenses [19-21] allows the system to perform long-range imaging without the need for mechanical refocusing.

Despite the various advantages of metalenses, challenges remain in addressing the increased chromatic and geometric aberrations introduced in lower invasiveness metalens-CFB systems. These aberrations tend to become more pronounced as the FOV increases, significantly degrading imaging performance and thus requiring dedicated correction methods. In recent years, data-driven deep learning approaches have been widely adopted for image restoration [22-25] such as deraining [26] and motion deblurring [27]. These methods do not rely exclusively on precise mathematical or physical models but instead leverage large datasets to train deep neural networks. Several studies have demonstrated that deep learning can enhance the image quality of metalenses. For instance, Dong *et. al.* employed MIRUnet to realize achromatic imaging using a single metalens [28], while Seo *et. al.* demonstrated an improved deep neural network (DNN) capable of reconstructing images from metalenses without dispersion engineering [29]. Notably, these metalenses exhibit relatively low NA of 0.298 and 0.2, which inherently reduces the complexity of deep learning restoration. In contrast, high-NA metalenses introduce strong aberrations and chromatic dispersion, requiring advanced neural networks for effective restoration.

Additionally, improving the fabrication quality of metalenses can help reduce chromatic dispersion and aberrations induced by fabrication errors. Several studies have employed materials such as metals or metal oxides as hard masks to realize high aspect ratio nanopillars [30, 31]. A common limitation among these approaches is the lack of sufficient sidewall verticality, which leads to phase delays deviating from the ideal values and ultimately deteriorates the imaging performance. This highlights the need to address the sidewall verticality of small cross section, high aspect ratio nanostructures through fabrication process optimization.

In this work, we present a metalens imaging system in the visible specifically designed for CFB. The system reduces the rigid tip length to 1 mm while expanding the FOV to 48.3° and extending the DOF beyond 125 mm. We design the metalens layout using a genetic algorithm (GA) with angularly consistent modulation transfer function (MTF) as the optimization target, while neglecting coupling between nanopillars to significantly reduce computational cost. To compensate for the aberrations and chromatic distortions caused by this approximation, we develop an MHCE process, which utilizes the metal hard mask not only as a pattern transfer layer but also as a sidewall compensation tool to achieve high aspect ratio nanopillars with vertical sidewalls. This process is compatible with standard CMOS processes. We anticipate that this metalens imaging platform will accelerate the integration of metalens-based optics into compact, ultra-minimally invasive CFB endoscopic systems.

## 2. Methods

### A. Overview

Fig. 1 illustrates the core components of the metalens-CFB system imaging experiment. A compact endoscopic system comprises a metalens and CFB, while the rear end includes an imaging module composed of an objective lens, tube lens and a CMOS camera. The "Metalens design" module depicts the genetic algorithm optimization process for determining the spatial arrangement of nanopillars. Each population consists of 80 candidate metalens phase profiles, which are used to compute MTFs across various incident angles and wavelengths. Based on the loss function derived from angular consistency criteria, the population undergoes iterative genetic operations, including crossover and mutation, to evolve toward an optimal design. In parallel, the MISRGAN module performs image restoration. The two modules are connected via the angularly consistent MTF.

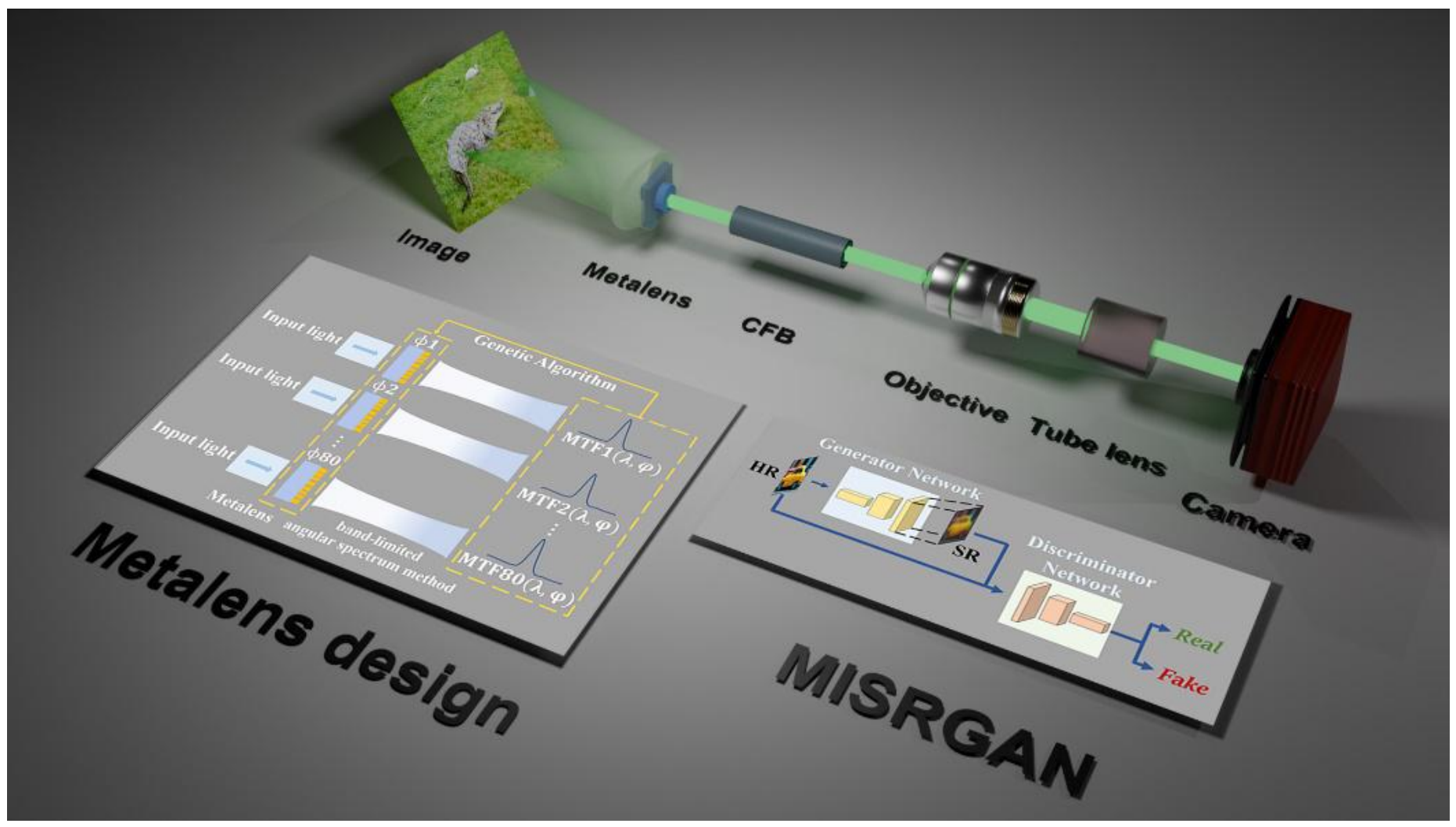

**Fig. 1.** The schematic diagram of the design, imaging and post-processing workflow of the metalens system. The upper optical path presents the complete Metalens-CFB imaging model, while the lower two boxes depict the design schematic of the metalens and the MISRGAN image restoration algorithm, respectively.

## B. Metalens design and fabrication

In the past few years, metalens design has primarily focused on micrometer-scale, where numerical simulation methods such as finite-difference time-domain (FDTD) [32] and finite element method (FEM) require manageable memory and computational requirements. The design process becomes significantly more resource-intensive with the growing interest in developing larger-aperture metalenses at millimeter to centimeter scales. For instance, simulating a 0.5 mm diameter metalens using FDTD requires terabytes of memory [30]. To address this issue, we adopt a simplified modeling approach that neglects the coupling effects between adjacent nanopillars and employs the band-limited angular spectrum method [33] to predict the metalens focusing characteristics. This method enables efficient 6-angle, 3-wavelength optimization using only ~2 GB of memory per iteration, with each iteration consuming 59.5 seconds. Although this approximation will introduce localized phase distortion and image aberrations, they can be effectively corrected in post-processing through our proposed MISRGAN. Benefiting from the strong nonlinear approximation capacity of deep neural networks, our method achieves marked improvements in peak signal-to-noise ratio (PSNR) and perceptual loss.

To achieve broadband imaging across the visible spectrum, the most rigorous approach is to optimize the metalens over multiple dense wavelengths, which will produce a heavy computational burden. Previous studies have demonstrated that optimizing solely for RGB wavelengths can yield satisfactory performance [34]. In this work, we select 606 nm, 511 nm, and 462 nm as the target wavelengths for optimization. The spatial distribution of the nanopillars is parameterized using a polynomial phase profile described in Equation (1), where r denotes the radial coordinate, λ represents the incident light wavelength and $a_k$ are the even-order coefficients to be optimized. After balancing accuracy and computational cost, the number of even-order terms is set to nine. Owing to the rotational symmetry of the metalens, all odd-order coefficients are set to zero. The phase delay difference Δϕ depends on the nanopillar dimension and the wavelength. The optimization uses a genetic algorithm [35], which performs a population-based global search to avoid a local optimum. This makes GA particularly suitable for the high-dimensional, nonlinear optimization of metalens phase profiles.

$$\phi(\mathrm{r},\lambda) = \sum_{k=0}^{9} a_k(\lambda = 511) r^{2k} + \Delta\phi(\lambda), \quad (1)$$

$$Loss = -\log\left(\prod_{i=1}^{3}\prod_{j=1}^{6}\mathrm{MTF}(\lambda_i,\theta_j)\right). \quad (2)$$

In the MISRGAN framework, the generator restores images by applying small convolutional kernels across the entire image. An angularly consistent MTF can improve the MISRGAN restoration performance. Therefore, we use MTF consistency as the loss function. The metalens optimization function is given by Equation (2), where the logarithmic term involves the product of MTF values computed at three design wavelengths and six incident angles (0°, 5°, 10°, 15°, 20°, and 25°). The MTF is the Fourier transform of the point spread function (PSF). The optimization range of the FOV is calculated based on the FOV of the CFB and the focal length of the metalens.

After optimization, the final metalens design consists of nanopillars with a uniform height of 600 nm and a maximum aspect ratio of ~6. Due to the great etch depth, conventional inductively coupled plasma (ICP) etching recipe often results in undercut profiles at the bottom of the nanopillars. Such structural changes directly affect the phase delay (see Appendix A), leading to divergence of the focal spot. To address this, we introduce a MHCE method, in which the RF power is increased and the fluorine-to-hydrogen (F/H) ratio in the etching gas is reduced during deep etching stages. This adjustment effectively suppresses ion beam divergence and mitigates fluorine-induced lateral etching. It should be noted that during the shallow etching stage before the adjustment, the metal mask experiences minimal consumption. However, after the conditions are adjusted, accelerated thinning of the metal mask occurs due to the intensified ion bombardment, particularly at the edges, necessitating close monitoring of mask degradation.

## 3. Results

The fabricated metalens is shown in Fig. 2(a) and (b), exhibiting complete structural integrity. See Appendix B for fabrication details. Compared to the single-step etching process [Fig. 2(d)], our improved MHCE method achieves nearly vertical sidewalls [Fig. 2(c)], which significantly enhances the phase accuracy, particularly for nanopillars with a smaller cross-section (see Appendix C). Using the MHCE process, we achieve vertically etched nanopillars with a minimum width of ~100 nm and an aspect ratio of ~6. As shown in Fig.2(e), the focal lengths of the metalens at the RGB wavelengths show an average deviation of approximately 0.13 mm; the discrepancy could be reduced by narrowing the spectral bandwidth of the LED light source (currently 20 nm). The minimal focal length deviation serves as a validation of the accuracy of the metalens design method. The corresponding PSFs and MTFs are shown in Fig. 2(f) and (g), with the experimental setup illustrated in Appendix D. The PSFs remain nearly the same within a 10° field angle, while noticeable tailing appears beyond 20°, indicating the off-axis aberrations. The MTFs (Fourier transforms of the PSFs) exhibit consistent profiles across all measured angles, with negligible wavelength dependence, demonstrating the metalens's broadband and wide-field imaging capabilities.

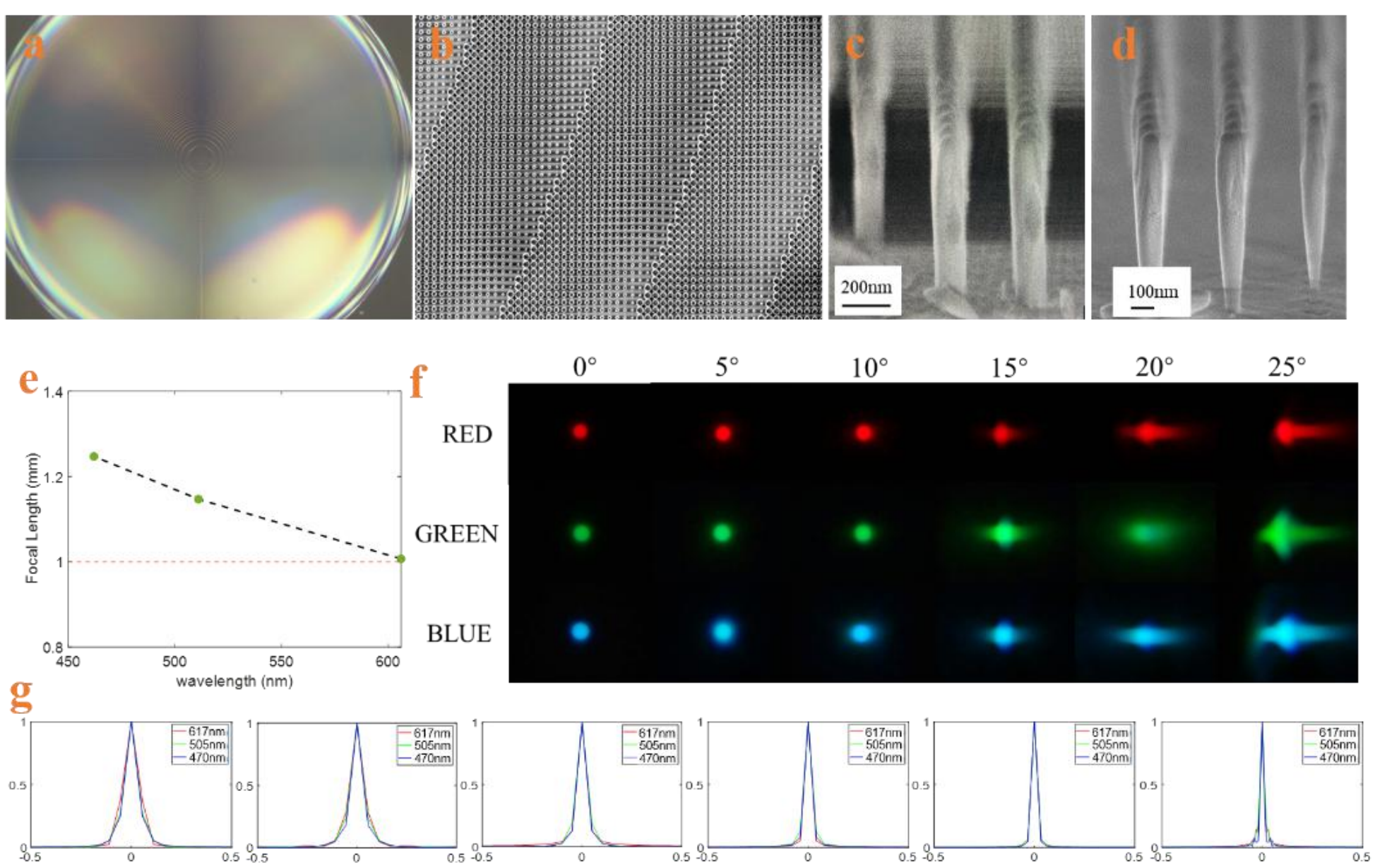

**Fig. 2.** (a) Metalens image captured by a COMS camera. The diameter is 1mm. (b) SEM image of the metalens. (c) Vertical sidewalls achieved using the MHCE method. (d) Tapered sidewalls resulting from a single-step etching process. (e) Focal lengths for RGB wavelengths, with the orange dashed line indicating the design value. (f) PSFs for RGB wavelengths across six incident angles. (g) Corresponding MTFs computed from the PSFs in (f). The MTFs are normalized for comparison.

.

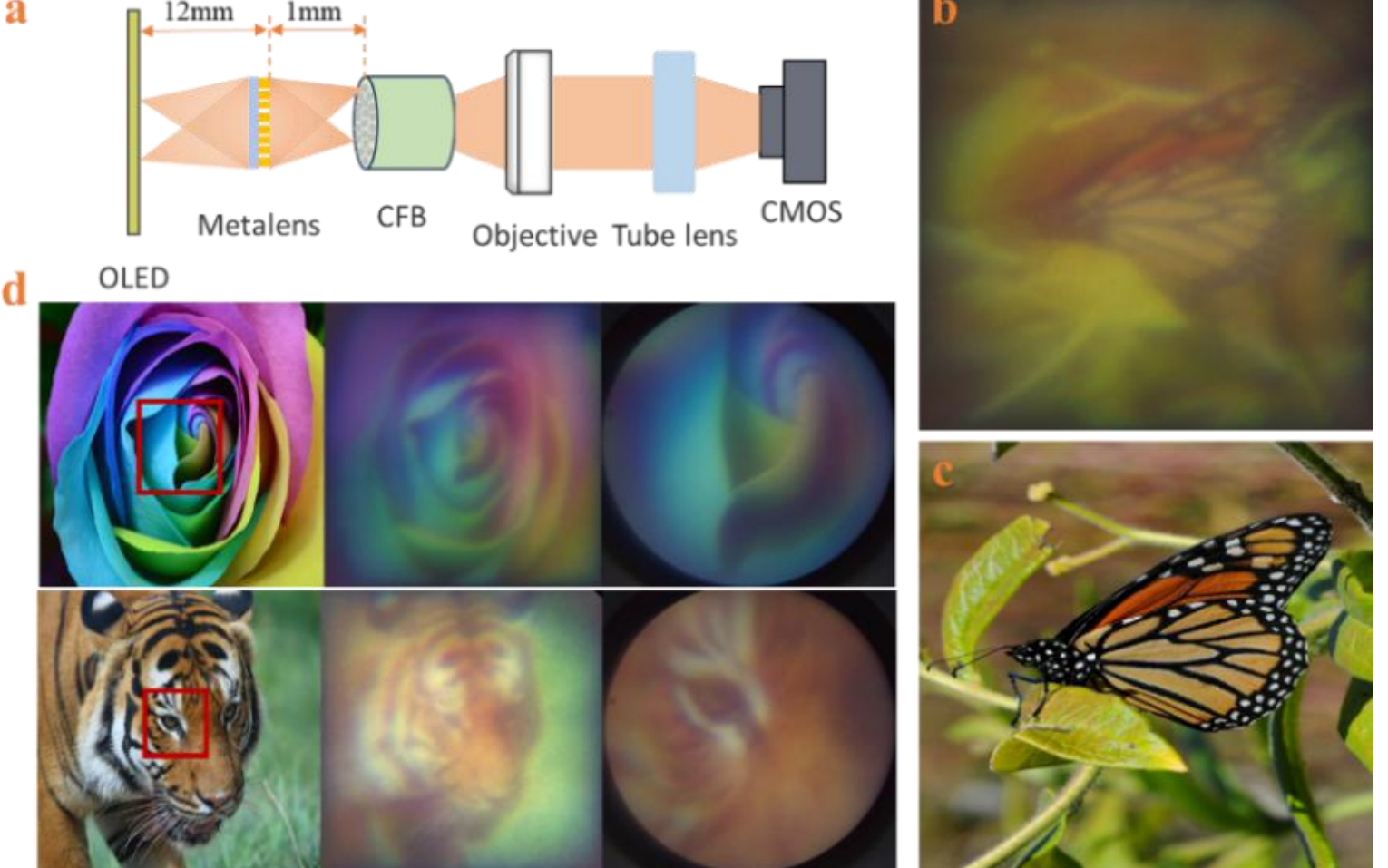

**Fig. 3.** (a) Schematic of the metalens-CFB imaging system, from left to right: OLED display, metalens, CFB, 10× objective lens, tube lens, and CMOS camera. (b) Image captured by CMOS with honeycomb artifacts. (c) Corresponding GT image of (b). (d) From left to right: GT image, Gaussian-blurred of the CMOS-captured image, and magnified view of the region enclosed by the red rectangular box in the GT image through system.

The designed metalens is integrated with a CFB to enable image acquisition at the distal end within a confined FOV. In the metalens-CFB imaging setup, the GT image is displayed on an OLED screen positioned 12 mm from the metalens. The metalens is placed 1 mm from the distal end of a commercial CFB fiber (Fujikura FIGH-850N) to capture the image. Geometrical optics calculations indicate that the effective full FOV is approximately 48.3°, within the optimized angular range of the metalens design. At the proximal end of the CFB, a 10× objective lens magnifies the transmitted image before it is captured by a CMOS camera [Fig. 3(a)]. The GT image projected on the OLED screen [Fig. 3(c)] is captured as Fig. 3(b), which exhibits characteristic honeycomb artifacts inherent to the CFB structure. These artifacts are effectively suppressed using a Gaussian filter with a kernel size of 15, as shown in the second column of Fig. 3(d). The average processing time per image is approximately 100 ms on a personal computer. The third column presents a magnified view of the region highlighted by the red rectangle in the GT image, providing detailed demonstration of the imaging capability of the metalens-CFB system.

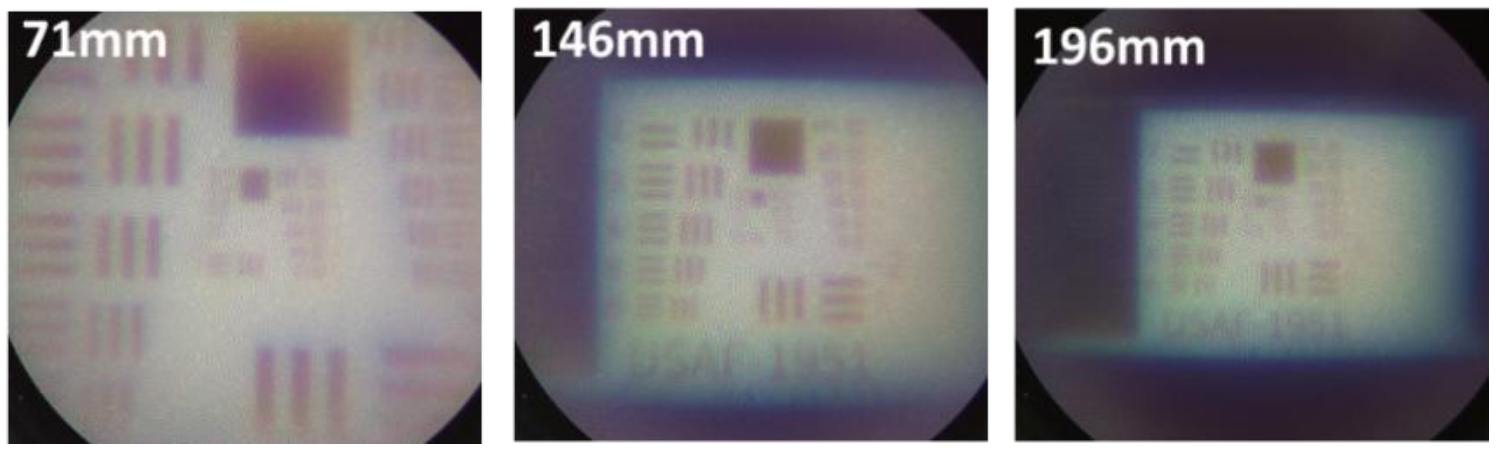

**Fig. 4.** The imagess of the USAF resolution target obtained by the metalens-CFB system at distances of 71 mm, 146 mm, and 196 mm.

We further imaged a USAF resolution target across a 125 mm DOF [Fig. 4], confirming that the system can resolve the outermost elements over an extended depth range. This large DOF is advantageous for endoscopic applications, enabling simultaneous clear visualization of tissues at different depths.

## 4. Discussion

Medical imaging is pivotal in modern clinical practice, providing essential guidance for disease diagnosis and treatment. Image reconstruction serves as a cornerstone of medical imaging. Its core objective is generating high-quality clinical images while minimizing cost and risk.

In this work, we designed a silicon nitride metalens specifically for CFB endoscope, achieving the least invasive metalens-CFB system reported to date, while achieving a wide FOV of 48.3°. Structurally, the system features a tip thickness of 1 mm, with the metalens oriented away from the external environment such that the wafer substrate acts as an inherent protective layer. A meaningful direction for future research is the integration of discrete components into a unified and manufacturable metalens-CFB system. Two major technical challenges must be overcome in this regard: dicing the metalens from the wafer and identifying a stable bonding material. Notably, if the refractive index of the bonding material is higher than that of the metalens substrate, the NA of the metalens can be enhanced without additional optimization, thereby further reducing the invasiveness of the system.

Here, we address chromatic dispersion and aberrations in high-NA metalens under air condition, achieving a 35% increase in PSNR and a 57.7% reduction in LPIPS. The key to the advancement lies in the synergistic integration of metalens design and deep learning–based image post-processing, rather than treating these components as independent stages. The mutual bridge between the two processes is the MTF. Furthermore, a genetic algorithm is employed to optimize the nanopillar distribution, effectively mitigating convergence to local minima and enhancing overall design robustness.

In previous studies on high-aspect-ratio metalens in the visible, the issue of sidewall non-verticality of nanostructures has rarely been addressed. Given that the feature dimensions of metalenses scale proportionally with the operational wavelength, shorter wavelengths require smaller cross-sectional dimensions, which exacerbate the risk of undercutting during fabrication of high-aspect-ratio nanostructures. To address this challenge, we developed the MHCE dry etching method, which enables vertical sidewall etching of nanopillar with small cross-sections (~100 nm) and high aspect ratios (~6). This approach significantly improves structural fidelity and enables precise phase control for visible-wavelength metalens performance. The proposed MHCE method enhances the fabrication precision of all silicon nitride-based metasurface devices, including high-quality (Q) factor resonators, vortex beam generators, and AR/VR applications. This improvement in fabrication quality directly contributes to enhanced overall device performance.

## APPENDIX A: Phase delay errors induced by non-vertical sidewalls

To simulate the impact of non-vertical sidewalls, we employ tapered structures to approximate actual undercuts and analyze the effect of varying undercut widths on phase delay. As shown in Fig. 5, larger undercut widths lead to greater phase delay errors. When the undercut width reaches 45nm, the duty cycle required to achieve the same phase delay differs by 0.14 between vertical and tapered structures, indicating a significant deviation.

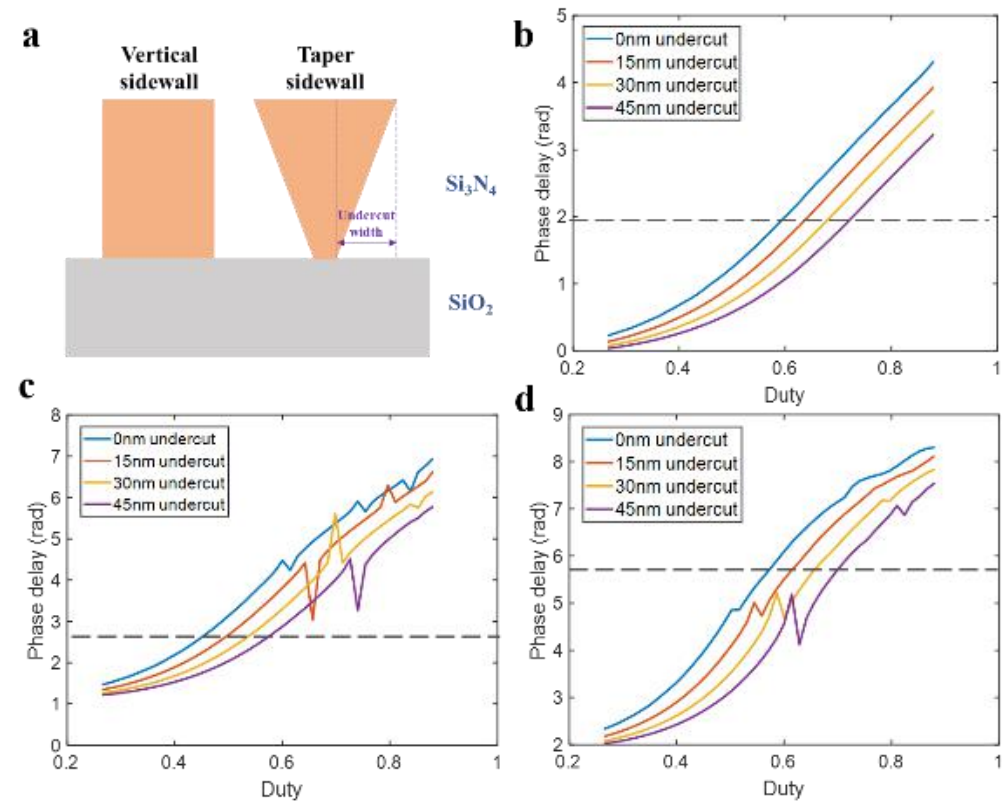

**Fig. 5.** (a) Schematic of vertical and tapered sidewalls for simulation. Phase delay errors under different undercut widths are simulated: (b) 606nm, (c) 511nm, and (d) 462nm wavelengths. The horizontal axis represents the duty cycle of the nanopillars.

## APPENDIX B: Sample fabrication

Beginning with a double side polished fused silica wafer, we deposit a 600 nm silicon nitride device layer using plasma-enhanced chemical vapor deposition (PECVD). We then spin coat with ARP-6200 electron beam resist and conductive polymer layer AP-PC5092.02 followed by exposure with an Elionix ELS-F125 electron-beam lithography system at 125 kV and 100 pA. After exposure, we remove the conductive layer with deionized water, and develop the resist using AR 600-546 developer. To define the etch mask, we evaporate 55 nm of aluminum and lift off via heated NMP, acetone, and isopropyl alcohol. We then etch the silicon nitride layer using $CHF_3$ and $SF_6$ gases with MHCE ICP processes. The gas ratios and RF powers in each step are different. Finally, we remove the aluminum mask using diluted hydrochloric acid. A graphical summary of the fabrication process of the metalens is shown in Fig. 6.

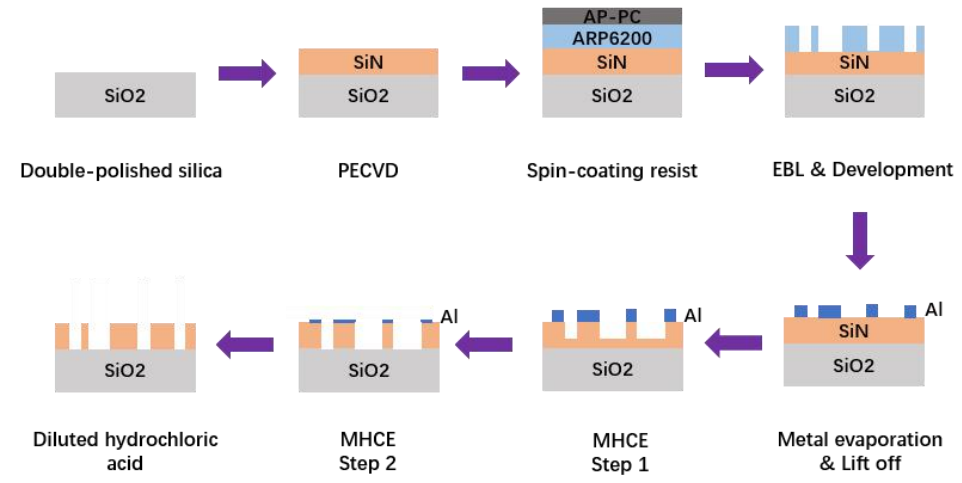

**Fig. 6.** Fabrication flow for the SiN metalens.

## APPENDIX C: Metal-Height Compensated Etching

To demonstrate the effectiveness of the MHCE process, we compare the results of MHCE, Step 1 condition of MHCE and Step 2 condition of MHCE, as shown in Fig. 7. Compared to Step 1, Step 2 employs a lower $SF_6$/ $CHF_3$ ratio and higher RF power. The sidewalls etched by the full MHCE process exhibit superior verticality compared to the other two conditions. Etching under Step 1 alone results in a highly vertical upper region [above the yellow line in Fig. 7(b)], while the lower region becomes tapered due to lateral fluorine etching and insufficient passivation. In contrast, Step 2 yields nearly vertical sidewalls in the lower region, but the upper portion broadens as a result of excessive passivation.

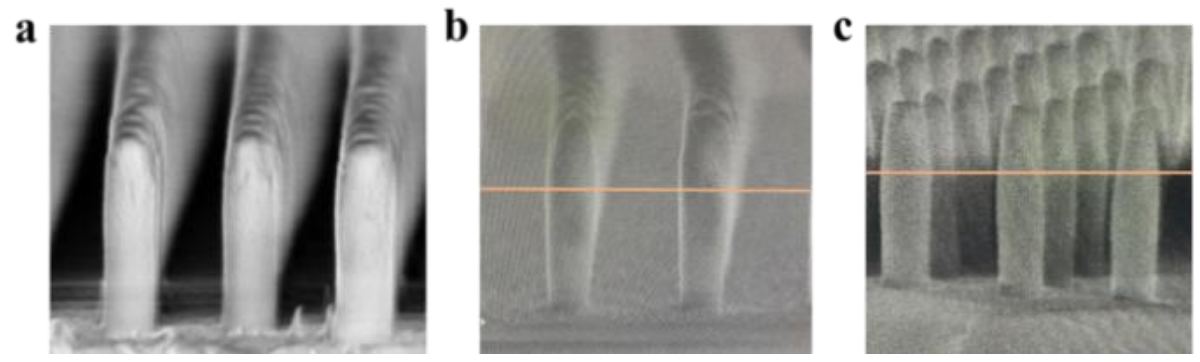

**Fig. 7.** Cross sectional SEM of the silicon nitride nanopillars etched by MHCE(a), MHCE step 1 condition(b), MHCE step 2 condition(c).

## APPENDIX D: Expermental setup for characterizing the focal plane

The illumination source is fiber-coupled LED (Thorlabs M470F4, M505F3, M617F2) followed by a beam collimator to generate collimated incident light. The metalens is mounted on a high-precision rotation stage (DMRP-1TA), enabling angular-dependent focal property characterization through controlled azimuthal rotation. A 20× objective lens coupled with a tube lens is positioned between the metalens and CMOS camera to magnify the focal spot image for detailed analysis. The expermental setup is shown in Fig. 8.

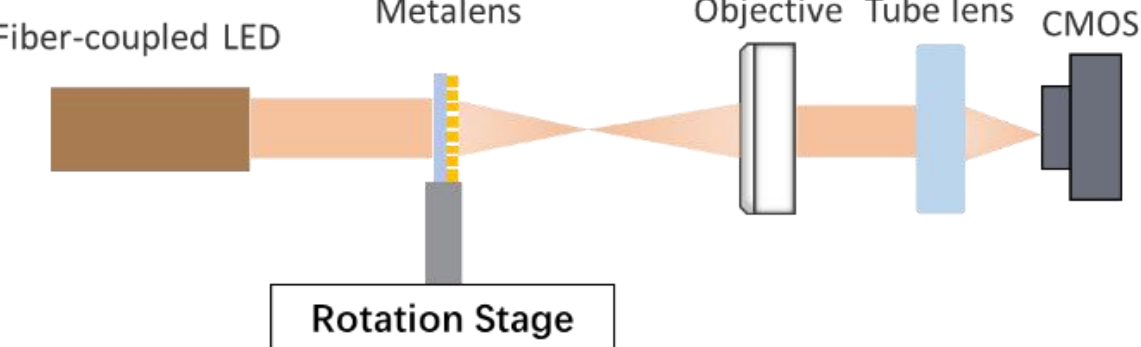

**Fig. 8.** Expermental setup for characterizing the focal plane. From left to right: a fiber-coupled LED, a metalens holder with a rotation stage, an objective lens, a tube lens, and a CMOS camera.

## Funding

## Acknowledgment

Computing resources are supported by High-performance Computing Platform of Peking University.

## Disclosures

The authors declare no conflicts of interest.

## Data availability

Data underlying the results presented in this paper are not publicly available at this time but may be obtained from the authors upon reasonable request.

## References

1. P. C. de Groen, "History of the endoscope," Proc. IEEE 105, 1987–1995 (2017).
2. L. E. Savastano, M. G. Muto, R. M. Starke, *et al*., "Diagnostic and interventional optical angioscopy in ex vivo carotid arteries," Oper. Neurosurg. 13, 36-46 (2017).
3. F. Litvack, W. J. Grundfest, J. S. Forrester, *et al*., "Angioscopic visualization of blood-vessel interior in animals and humans," Clin. Cardiol. 8, 65-70 (1985).
4. P. Z. McVeigh, T. Moloney, B. C. Wilson, *et al.*, "High-resolution scanning fiber angioscopy as an adjuvant to fluoroscopy during endovascular interventions," J. Endovasc. Ther. 25, 617-623 (2018).
5. K. W. Shim, E. K. Park, D. S. Kim, *et al.*, "Neuroendoscopy: current and future perspectives," J. Korean Neurosurg. Soc. 60, 322-326 (2017).
6. H. Arishima, K. Sato, H. Kikuta, *et al.*, "Spinal endoscopy combined with selective CT myelography for dural closure of the spinal dural defect with superficial siderosis: technical note," J. Neurosurg. Spine 28, 96-102 (2018).
7. N. Xie, Z. Zhou, J. E. Fröch, *et al.*, "Inverse-designed large field-of-view polychromatic metalens for tri-color scanning fiber endoscopy," Commun. Eng. 4, 53 (2025).
8. M. Cui, Z. Cheng, Y. Li, *et al.*, "Integrated approach for large volume imaging through GRIN lenses," Proc. SPIE PC13303, PC1330309 (2025).
9. Shanker, A., Fröch, J.E., Mukherjee, S, et al. "Quantitative phase imaging endoscopy with a metalens", Light Sci Appl 13, 305 (2024).
10. Y. Yang, E. Lee, Y. Park, *et al.*, "The Road to Commercializing Optical Metasurfaces: Current Challenges and Future Directions," ACS Nano 19, 3008-3018 (2025).
11. H. Liang, A. Martins, B.-H. V. Borges, *et al.*, "High performance metalenses: numerical aperture, aberrations, chromaticity, and trade-offs," Optica 6, 1461-1470 (2019).
12. W. T. Chen, A. Y. Zhu, and F. Capasso, "Flat optics with dispersion-engineered metasurfaces," Nat. Rev. Mater. 5, 604-620 (2020).
13. M. Pan, Y. Fu, M. Zheng, *et al.*, "Dielectric metalens for miniaturized imaging systems: progress and challenges," Light Sci. Appl. 11, 195 (2022).
14. A. Arbabi and A. Faraon, "Advances in optical metalenses," Nat. Photonics 17, 16-25 (2023).
15. M. Zhao, M. K. Chen, Z. P. Zhuang, *et al.*, "Phase characterisation of metalenses," Light Sci. Appl. 10, 52 (2021).
16. A. Arbabi, Y. Horie, A. Ball, *et al.*, "Subwavelength-thick lenses with high numerical apertures and large efficiency based on high-contrast transmitarrays," Nat. Commun. 6, 7069 (2015).
17. W. T. Chen, A. Y. Zhu, V. Sanjeev, *et al.*, "A broadband achromatic metalens for focusing and imaging in the visible," Nat. Nanotechnol. 13, 220-226 (2018).
18. J. Wu, T. Wang, O. Uckermann, *et al.*, "Learned end-to-end high-resolution lensless fiber imaging towards real-time cancer diagnosis," Sci. Rep. 12, 18846 (2022).
19. C. Chen, W. Song, J. W. Chen, *et al.*, "Spectral tomographic imaging with aplanatic metalens," Light Sci. Appl. 8, 99 (2019).
20. B. Xu, H. Li, S. Gao, *et al.*, "Metalens-integrated compact imaging devices for wide-field microscopy," Adv. Photonics 2, 066004 (2020).
21. W. T. Chen, A. Y. Zhu, F. Capasso, *et al.*, "Immersion Meta-Lenses at Visible Wavelengths for Nanoscale Imaging," Nano Lett. 17, 3188-3194 (2017).
22. Z. Qin, Q. Zeng, Y. Zong, *et al.*, "Image inpainting based on deep learning: A review," Displays 69, 102028 (2021).
23. C. Ledig, L. Theis, F. Huszar, *et al.*, "Photo-Realistic Single Image Super-Resolution Using a Generative Adversarial Network," 2017 IEEE Conference on Computer Vision and Pattern Recognition (CVPR), 105-114 (2017).
24. C. Dong, C. C. Loy, K. He, *et al.*, "Image Super-Resolution Using Deep Convolutional Networks," IEEE Trans. Pattern Anal. Mach. Intell. 38, 295-307 (2016).
25. P. Dhariwal and A. Nichol, "Diffusion Models Beat GANs on Image Synthesis," arXiv:2105.05233 (2021).
26. J. H. Kim, J. Y. Sim, and C. S. Kim, "Video deraining and desnowing using temporal correlation and low-rank matrix completion," IEEE Trans. Image Process. 24, 2658-2670 (2015).

27. O. Kupyn, V. Budzan, M. Mykhailych, *et al.*, "DeblurGAN: Blind Motion Deblurring Using Conditional Adversarial Networks," 2018 IEEE/CVF Conference on Computer Vision and Pattern Recognition, 8183-8192 (2018).
28. Y. Dong, W. T. Chen, A. Y. Zhu, *et al.*, "Achromatic Single Metalens Imaging via Deep Neural Network," ACS Photonics 11, 1645-1656 (2024).
29. J. Seo, J. Jo, J. Kim, *et al.*, "Deep-learning-driven end-to-end metalens imaging," Adv. Photonics 6, 066002 (2024).
30. E. Tseng, S. Colburn, J. Whitehead, *et al.*, "Neural nano-optics for high-quality thin lens imaging," Nat. Commun. 12, 6493 (2021).
31. J. E. Fröch, L. Huang, Q. A. Tanguy, *et al.*, "Real time full-color imaging in a Meta-optical fiber endoscope," eLight 3, 13 (2023).
32. M. Mansouree, A. McClung, A. Arbabi, *et al.*, "Multifunctional 2.5D metastructures enabled by adjoint optimization," Optica 7, 77-84 (2020).
33. K. Matsushima and T. Shimobaba, "Band-Limited Angular Spectrum Method for Numerical Simulation of Free-Space Propagation in Far and Near Fields," Opt. Express 17, 19662-19673 (2009).
34. Z. Li, M. K. Chen, Y. H. Fu, *et al.*, "Meta-optics achieves RGB-achromatic focusing for virtual reality," Sci. Adv. 7, eabe4458 (2021).
35. I. Harvey, "The Microbial Genetic Algorithm," Fourth European Conference on Artificial Life, 434-443 (2009).
36. Y. Zhang, Y. Tian, and Y. Kong, *et al.*, "Residual Dense Network for Image Restoration," IEEE Trans. Pattern Anal. Mach. Intell. 43, 2480-2495 (2021).
37. B. Lim, S. Son, and H. Kim, "Enhanced Deep Residual Networks for Single Image Super-Resolution," 2017 IEEE Conference on Computer Vision and Pattern Recognition Workshops (CVPRW), 1132-1140 (2017).
38. Xiao, X., Zhao, Y., Ye, X. *et al.* Large-scale achromatic flat lens by light frequency-domain coherence optimization. Light Sci Appl 11, 323 (2022).